# Automated Quantification of the Impact of the Wood-decay fungus *Physisporinus vitreus* on the Cell Wall Structure of Norway spruce by Tomographic Microscopy

M. J. Fuhr[a,b], C. Stührk[a,b], B. Münch[c], F. W. M. R. Schwarze[b] and M. Schubert[b]

[a]*ETH Zurich, Institute for Building Materials, Computational Physics for Engineering Materials, Schafmattstrasse 6, HIF E18, CH-8093 Zurich, Switzerland*

[b]*EMPA, Swiss Federal Laboratories for Materials Science and Technology, Wood Laboratory, Group of Wood Protection and Biotechnology, Lerchenfeldstrasse 5, CH-9014 St Gallen, Switzerland*

[c]*EMPA, Swiss Federal Laboratories for Materials Science and Technology, Group Concrete Technology, CH-8600 Dübendorf, Switzerland*

Corresponding author: M.J. Fuhr. Phone: +41 44 633 7153

Email addresses:   MF: mfuhr@ethz.ch
                   CS: Chris.Stührk@empa.ch
                   BM: Beat.Muench@empa.ch
                   FS: Francis.Schwarze@empa.ch
                   MS: Mark.Schubert@empa.ch

22    4 figures

23    4372 words

24

25    **Abstract**

26    Wood-decay fungi decompose their substrate by extracellular, degradative enzymes and play an

27    important role in natural ecosystems by recycling carbon and minerals fixed in plants. Thereby,

28    they cause significant damage to the wood structure and limit the use of wood as building

29    material. Besides their role as biodeteriorators wood-decay fungi can be used for

30    biotechnological purposes, e.g. the white-rot fungus *Physisporinus vitreus* for improving the

31    uptake of preservatives and wood-modification substances of refractory wood. Therefore, the

32    visualization and the quantification of microscopic decay patterns are important for the study of

33    the impact of wood-decay fungi in general, as well as for wood-decay fungi and microorganisms

34    with possible applications in biotechnology. In the present work, we developed a method for the

35    automated localization and quantification of microscopic cell wall elements (CWE) of Norway

36    spruce wood such as bordered pits, intrinsic defects, hyphae or alterations induced by *P. vitreus*

37    using high resolution X-ray computed tomographic microscopy. In addition to classical

38    destructive wood anatomical methods such as light or laser scanning microscopy, our method

39    allows for the first time to compute the properties (e.g. area, orientation and size-distribution) of

40    CWE of the tracheids in a sample. This is essential for modeling the influence of microscopic

41    CWE to macroscopic properties such as wood strength and permeability.

42

46

47   **1. Introduction**

48   Wood-decay fungi degrade their substrate, a complex anisotropic material featuring several

49   hierarchical levels of organization from macroscopic (e.g. growth ring) over the mesoscopic (e.g.

50   set of wood cells) down to the microscopic and nanoscopic scale (e.g. wood cells and fibrils), by

51   extracellular, degradative enzymes. They play an important role in natural ecosystems by

52   recycling carbon and minerals fixed in plants. Thereby, they cause a significant damage to the

53   wood structure and limit the possible use of wood as building material. Besides their role as

54   biodeteriorators wood-decay fungi can be used for biotechnological applications.

55        Recent investigations have shown that wood-decay fungi have many valuable

56   biotechnological purposes in the pure and applied wood sciences (Mai et al., 2004; Messner et

57   al., 2002; Schwarze, 2008). Alterations in the cell wall structure and/or the distribution of the cell

58   wall constituents are reflected in the plasticity of the wood degradation modes of different fungi

59   (Deflorio et al., 2005; Schwarze, 2008). The specificity of their enzymes and the mild conditions

60   under which degradation proceeds potentially make them suitable agents for wood modification

61   such as biopulping, bioremendiation or bioincising (Majcherczyk and Hüttermann, 1988;

62   Messner et al., 2002; Schwarze, 2009).

63        The biotechnological process of bioincising is a promising approach for improving the

64   uptake of preservatives and wood-modification substances by refractory wood due to

65   degradation of bordered pits by the white-rot fungus *Physisporinus vitreus* (Pers.: Fr.) P. Karst.

66   (Schwarze and Landmesser, 2000; Schwarze et al., 2006). Furthermore because of its exceptional

pattern of degradation, *P. vitreus* is successfully used to improve the acoustic properties of the tonewood of Norway spruce wood (*Picea abies* L.) used for music instruments by selectively delignifying the secondary walls without affecting the middle lamellae, even at advanced stages (Schwarze et al., 2008; Spycher et al., 2008). However, a successful up-scaling of biotechnological processes in which *P. vitreus* is used to improve substrate properties requires a set of investigations for the identification and detection of important growth parameters (Schubert and Schwarze, 2010; Schubert et al., 2010; Schubert et al., 2009) and the elucidation of the wood - fungus interactions (Lehringer et al., 2010). Hence, the visualization and quantification of microscopic decay patterns are of interest for the study of wood-decay fungi in general as well as for wood-decay fungi and microorganisms with possible applications for biotechnology. However, the quantification of microscopic cell wall alterations (e.g. their size-distribution in space and time) is difficult, because of the opacity of wood and the heterogeneity of its structure.

Classical destructive methods for analyzing wood-decay fungi include light and electron microscopy of thin microtome sections. The advantages of the latter methods are the elucidation and interpretation of wood-decay patterns with the help of specific staining techniques (Schwarze, 2007). However, classical microscopy techniques only provide two-dimensional (2D) information and do not allow the quantification of alterations in the cell wall structure. Yet, the growth of fungi in wood is a complex three-dimensional (3D) process due to the diverse alignment of wood cells and the distribution of nutrients. In order to successfully model the growth and impact of wood-decay fungi, more quantitative information on the distribution of fungal activity at the microscopic level in space and time is required (Fuhr et al., 2010).

At the microscopic scale, non-destructive techniques based on X-ray computed tomographic (XCT) microscopy have been mostly used in wood research for 3D investigation of mycelial expansion and the impact of wood-decay fungi. In XCT, the beam attenuation is acquired either by absorption or by scattering, according to the atomic number of the constituents at each volume element (voxel). The 2D projections are subsequently reconstructed into a 3D attenuation map. McGovern (McGovern et al., 2010) measured the mass loss of wood specimens using XCT with a voxel size of approximately 0.34 mm$^3$. Illman and Dowd (1999) and Van den Bulcke et al. (2009) analyzed the density and structure of incubated wood applying high-resolution XCT microscopy. Van den Bulcke et al. (2008) identified single hypha of *Aureobasidium pullulans* with a diameter of approximately 10 µm in wood using high-resolution XCT microscopy. However an automated separation of fungus and wood was not accomplished. It seems that confocal laser scanning microscopy (CLSM) is the appropriate technique to analyze the 3D structure of mycelium in wood, because of the capacity to separate fluorescence-stained fungi from the wood substrate during measurement (Dickson and Kolesik, 1999; Stührk et al., 2010). However, the penetration depth of the laser light into wood is limited and the cell wall damages are not clearly visible, because of the weak autofluorescence of wood. Therefore, we suggest the use of high-resolution XCT microscopy for analyzing the impact of wood-decay fungi.

To the best of our knowledge, the present work describes for the first time a computer based automated procedure for the localization and quantification of cell wall elements (CWE) such as bordered pits, intrinsic defects and alterations induced by *P. vitreus* by means of high-resolution XCT microscopy. The quantitative information arising from this procedure allows to e.g. analyze the distribution of the fungal activity of *P. vitreus* in the late- and early wood of

Norway spruce depending on the incubation conditions, which is essential for the successful manufacture of fungal modified wood.

## 2. Materials and methods

2.1 Wood and fungus

We use defect-free heartwood wood from a Norway spruce tree (*Picea abies* [L.] Karst.), grown in Switzerland. The three alignments of wood cells are longitudinal (parallel to the fibre), radial (perpendicular to the fibre) and tangential (parallel to the growth rings). There are mainly two types of cells in softwoods, tracheids and rays. The cell walls of tracheids consist of several layers denoted as secondary wall $S_1$, $S_2$, $S_3$ and primary wall (PW) from the lumen (i.e. voids within the cells) to the middle lamella (ML) forming the border of two adjacent tracheids. In order to transport water and nutrients in longitudinal and radial direction, the cell lumina are connected via bordered and simple pits.

Specimens with dimensions of approximately 400 µm (radial) $\times$ 10 mm (tangential) $\times$ 6 mm (longitudinal) were produced with a microtome. All specimen faces, except the radial ones, were subsequently coated by brushing (Nuvovern ACR Emaillack, Walter Mäder AG, Killwangen, Switzerland). After 24h, the procedure was repeated to guarantee a solid sealing. Subsequently, the specimens were conditioned for two weeks at 22°C and 50% relative humidity (RH). Thereafter, the specimens were sterilized with hot steam (121°C, 20 min and 2 bar) and placed on a 'feeder block' of Scots pines (*Pinus sylvestris*) previously colonized with the white-rot basidiomycete *Physisporinus vitreus* Empa strain 642. Specimens were incubated under sterile conditions for seven weeks at 22°C and 70% RH. After incubation, the specimens were cut into elongated wood prisms of approximately 400 µm (radial) × 400 mm (tangential) × 6 mm (longitudinal) using a microtome.

For the tomographic experiments, the samples were glued onto cylindrical sample holders using double-side adhesive tape, the longitudinal axis of the sample being located at the rotation axis of the tomographic stage.

2.2 Synchrotron Tomographic Microscopy

Synchrotron radiation facilities provide photon beams of energy densities that outrange conventional X-ray sources by orders of magnitude. Among other benefits, the high brilliance and brightness of synchrotron based X-rays enable tomographic microscopy at sub-micrometer scale.

In the present study, tomographic experiments were performed at the TOMCAT beam line (Tomographic Microscopy and Coherent Radiology experiments) at the synchrotron radiation facility Swiss Light Source (SLS) at the Paul Scherrer Institute (PSI) in Villigen (Switzerland). The TOMCAT beam line operates in both, absorption and phase-contrast mode. Phase-contrast tomography analyzes the Zernike phase-contrasts of the X-ray beam induced by refraction (Neuhäusler et al., 2003) and is preferable for materials with low absorption contrast such as wood. Trtik (Trtik et al., 2007) and Mannes (Mannes et al., 2010) demonstrated the use of phase-contrast tomography for the analysis of 3D structures in Norway spruce wood down to the microscopic level.

In order to minimize dehydration of the specimen during measurements a climatic chamber to control the air humidity at 95% RH (Derome et al., 2010) was used. The temperature remained constant during measurements at 25°C. For each specimen, a set of 1501 projections over 180° was acquired with a photon energy of 9.9 and 20.2 keV for absorption or phase-contrast mode, respectively. The X-rays were converted into visible light by a YAG:Ce 20 µm scintillator and projected to a charge coupled device (CCD) featuring a resolution of 2048 × 2048 pixels and a dynamic range of 14-bit. The nominal edge length of the cubic voxels was 0.37 µm by using an optical objective with the magnification (20×) and a field of view of 0.75 × 0.75 mm. The total scanning time was approximately 15 minutes for both, absorption and phase-contrast mode. Stampanoni et al. (2006) provides further technical specifications for TOMCAT.

The reconstruction of the original projections into a stack of 2048 transverse sections termed tomograms was based on Filtered-Back-Projection by using the Parzen filter supporting noise suppression. The tomograms are 16-bit gray-level TIFF images. The projection values were initially corrected with dark- and flat-field images and the attenuation values thereof were obtained by Lambert-inversion. Stripe artifacts originating from defective detector pixels were eliminated (Munch et al., 2009) and centering artifacts remedied.

2.3. Analysis of tracheid cell wall elements

Fig. 1 illustrates the cell wall analysis process and Fig. 2 presents the core algorithm. The original data, a stack of $n$ subsequent tomograms of the specimen is displayed in Fig. 1a, the data processing yielding the CWE in Figs. 1b - d. Initially, the ROI of each tracheid is manually identified in the 3D tomograms. The 3D tracheid objects are subsequently mapped into 2D by applying a cylindrical projection, each point representing the mean attenuation of the cell wall voxels at an angle $\alpha$, referring to the centre of gravity of the tracheid. The resulting tracheidal 2D-map is a gray-level image of size $n \times s$, where $s \sim 2\cdot\pi/\alpha$, in which regions of low attenuation values (i.e. CWE) are clearly visible. After segmentation, the distribution of the CWE size and the orientation was determined.

First, a region of interest (ROI) was selected and transformed into a binary mask separating air (black pixels) and wood material (white pixels). Segmentation was based on gray-level thresholding using Otsu's method (Otsu, 1979). Fig. 1b shows the original tomograms after applying a morphological closing operation by using a spherical structuring element (SE) with the radius of 3 pixels, in order to remove small objects mainly originating from noise (Gonzalez et al., 2009). The resulting filtered mask is the basis for the segmentation and mapping of the tracheids.

Tracheids were segmented by constructing the watersheds between adjacent and closed lumina (Meyer, 1994). Since the lumina might not be closed because of CWE (arrow in Fig. 3a), a morphological closing operation using a SE of $3 \times 3 \times 150$ pixels was applied (Fig. 1c) prior to watershed construction. Subsequently, each spanning cluster is labeled as shown by the colors in Fig. 1c. The obtained label mask of each tomogram makes the segmentation of a tracheid in the filtered mask possible and therefore the construction of a 3D tracheid mask, which is finally used to select and map the cell wall voxels to a tracheidal 2D-map by using a cylindrical projection. Fig. 1d illustrates this process for one of the tomograms.

## 3 Results and Discussion

In order to demonstrate the potential of the method we analyzed single tracheids of a ROI with the size of $125 \times 125 \times 400$ pixels using synchrotron XCT in absorption (Fig. 3a) and phase-contrast mode (Fig. 3b) visualized by isosurfaces in Figs. 3a and b respectively. CWE such as bordered pits (P), intrinsic defects or cell wall alterations induced by *P. vitreus* (F, L) were

202  marked. The tracheidal 2D-map of this tracheid is shown in Fig. 4a. The CWE are clearly visible
203  and their segmentation was possible (Fig. 4b). Fig. 4c shows the histogram of the CWE areas.
204  Their mean attenuation can be interpreted as a measure of cell wall damage as shown in the inset.
205  The results revealed the tracheid's lateral surface of approximately 9800 $\mu m^2$, had a total number
206  of CWE of 18 with an area of approximately 285 $\mu m^2$, which relates to approximately 3% of the
207  tracheid's lateral surface. The largest and the smallest CWE had areas of 72 $\mu m^2$ and 0.05 $\mu m^2$,
208  respectively. Most of the CWE were smaller than 40 $\mu m^2$. In Addition, most CWE occurred in
209  the tangential cell walls (Fig. 4a and b) and different shapes of CWE were recorded in tangential
210  and radial cell walls. Larger CWE showed a lower mean attenuation than smaller CWE.
211  The detection and computing of the CWE revealed clusters of pixels with a very low
212  attenuation as illustrated in the tracheidal 2D-map (Fig. 3b and Fig. 4a). However, the shape of
213  the cell wall and of the 'holes' strongly depends on the constant value for the isosurfaces (Figs.
214  3a and b) and the binary mask of the tracheidal 2D-map (Fig. 4a). Thus, for future measurements
215  it is necessary to compare the tracheids before and after fungal exposure in order to identify
216  alterations of the cell wall accurately. Therefore, the scanning procedure may be time
217  consuming, but recent developments make laboratory-based phase-contrast XCT microscopy
218  available (Mayo et al., 2010).
219  Since tracheids exhibit a complex 3D shape the presented cylindrical projection distorts
220  the cell wall and an elliptic cylindrical map projection might be more adequate. Furthermore,
221  there is more noise in the absorption than in phase-contrast based tomograms, which makes
222  analysis more difficult. Therefore, we suggest using phase-contrast based tomographic
223  microscopy. Nevertheless, the analysis shows pits and cell wall alterations that might be induced
224  by fungal activity, because the pattern of the damages were similar to those found in semi-thin light microscopy sections of incubated wood samples by Lehringer et al. (2010). We found that most of the cell wall alterations were located in the vicinity of bordered pits, and that the size-distribution in Fig. 4c shows a concentration of pixels with low attenuation to large CWE's such as cell wall alterations. This results corresponds with the findings of Lehringer et al. (2010) and according to his classification system we are able to classify the cell wall and bordered pits (indicated by L in Fig. 3a and b) as 'strongly degraded'.

Despite the inherent structure of wood, fungi degrade woody tissues, and decay types fall into three categories according to their mode of degradation of the woody cell walls. Traditionally, wood decomposition by fungi is usually classified into three categories based on micro-morphological and chemical characteristics of decay, resulting in different patterns of degradation of the cell wall: soft rot, brown rot and white rot, the latter subdivided into simultaneous rot and selective delignification as caused by e.g. *P. vitreus* (Schwarze 2008). Finally, the presented method has the potential to identify and quantify those cell wall alterations caused by different decay types and additionally other objects within wood such as bordered pits, intrinsic defects or hyphae by comparing the wood sample before and after fungal incubation.

**4 Conclusion**

We presented a method to analyze and quantify microscopic cell wall elements (CWE) such as pits, intrinsic defects and cell wall alterations induced by *Physisporinus vitreus*. Our analysis focused on Norway spruce tracheids degraded by the white-rot fungus *P. vitreus*. The CWE were clearly visible and it was possible to segment and determine the distribution of the CWE size and orientation.

We found that the most of the cell wall alterations were located in the vicinity of the bordered pits and the computed size-distribution shows a concentration of pixels with low attenuation to large CWE such as cell wall alterations. However, in addition to this classical wood anatomical method, for the first time our approach allowed to compute the properties (e.g. area, orientation and size-distribution) of cell wall elements of each tracheid of a specimen, which is essential for linking the influence of microscopic cell wall elements to macroscopic system properties such as wood strength or permeability.

Therefore, in the future we will systematically measure the fungal activity of *P. vitreus* in Norway spruce samples for different incubation periods and model the evolution of its impact to the cell wall structure. The obtained models are essential to simulate the permeability changes of infected wood in order to optimize the choice of pellet concentration and reaction times that are required to induce a certain degree of wood permeability by *P. vitreus*. Furthermore, the presented method facilitates the development and calibration of mathematical models to optimize the impact of wood decay-fungi for biotechnological applications in pure and applied wood sciences.

**Acknowledgments**
We acknowledge contributions and support of (in alphabetical order), Francois Gaignat, Hans Herrmann, Christian Lehringer, David Mannes, Peter Niemz, Pavel Trtik, and Falk Wittel. We thank to Masuru Abuku and Frederica Marone for their assistance during the measurements and Dominique Derome for supplying the climatic chamber. The authors express their gratitude to the Swiss National Foundation (SNF) No. 205321-121701 for its financial support.

**Figure captions**

Figure 1: Schematic workflow from (a) tomographic experiment to (b-d) quantification of cell wall elements. (a) Acquisition of original data, a stack of *n* subsequent tomograms, and manual identification of a ROI. (b) Segmentation of the tomograms to a binary mask and removing of artifacts. (c) Closing of the cell lumina in the filtered mask in order to obtain a label mask and a watershed mask to identify the pixels corresponding to a specific tracheid. (d) Construction of the tracheid mask and mapping procedure mapping. The 3D tracheid object is subsequently mapped into 2D by using a cylindrical projection, where each point represents the mean attenuation of the cell wall voxels at an angle $\alpha$, referring to the centre of gravity of the tracheid. Based on the resulting tracheidal 2D-map, which is a gray-level image of size $n \times s$, it is possible to segment and determine the distribution of the CWE size and orientation.

Figure 2: Core algorithm for computing the size-distribution of cell wall elements of a tracheid.

Figure 3: ROI with a size of $125 \times 125 \times 400$ pixels of a specimen incubated with *P. vitreus* for 8 weeks. The sample was measured by using the synchrotron tomographic microscopy in (a) absorption and (b) phase-contrast mode and visualized by isosurfaces. The clusters of pixels with a very low attenuation indicate cell wall elements (CWE) such as pits (P), intrinsic defects or cell wall alterations may induced by *P. vitreus* (F,L).

Figure 4: Quantification of cell wall elements (CWE). (a) Dark colors in the tracheidal 2D-map correspond to a low attenuation of the beam and therefore to a low density of the cell wall. The pits and cell wall alterations induced by the white-rot fungus *P. vitreus* are clearly visible. (b) Identification of the CWE by construction a threshold based binary mask of the tracheidal 2D-map. The tomogram of Fig. 1 corresponds to the marked row (dashed line). (c) An automated

segmentation of CWE allows analyzing e.g. their size-distribution. The inset shows the area of the CWE plotted against their mean attenuation.

**Figure s (original size)**

Dieser Text ist in Times New Roman 12
Dieser Text ist in Times New Roman 10

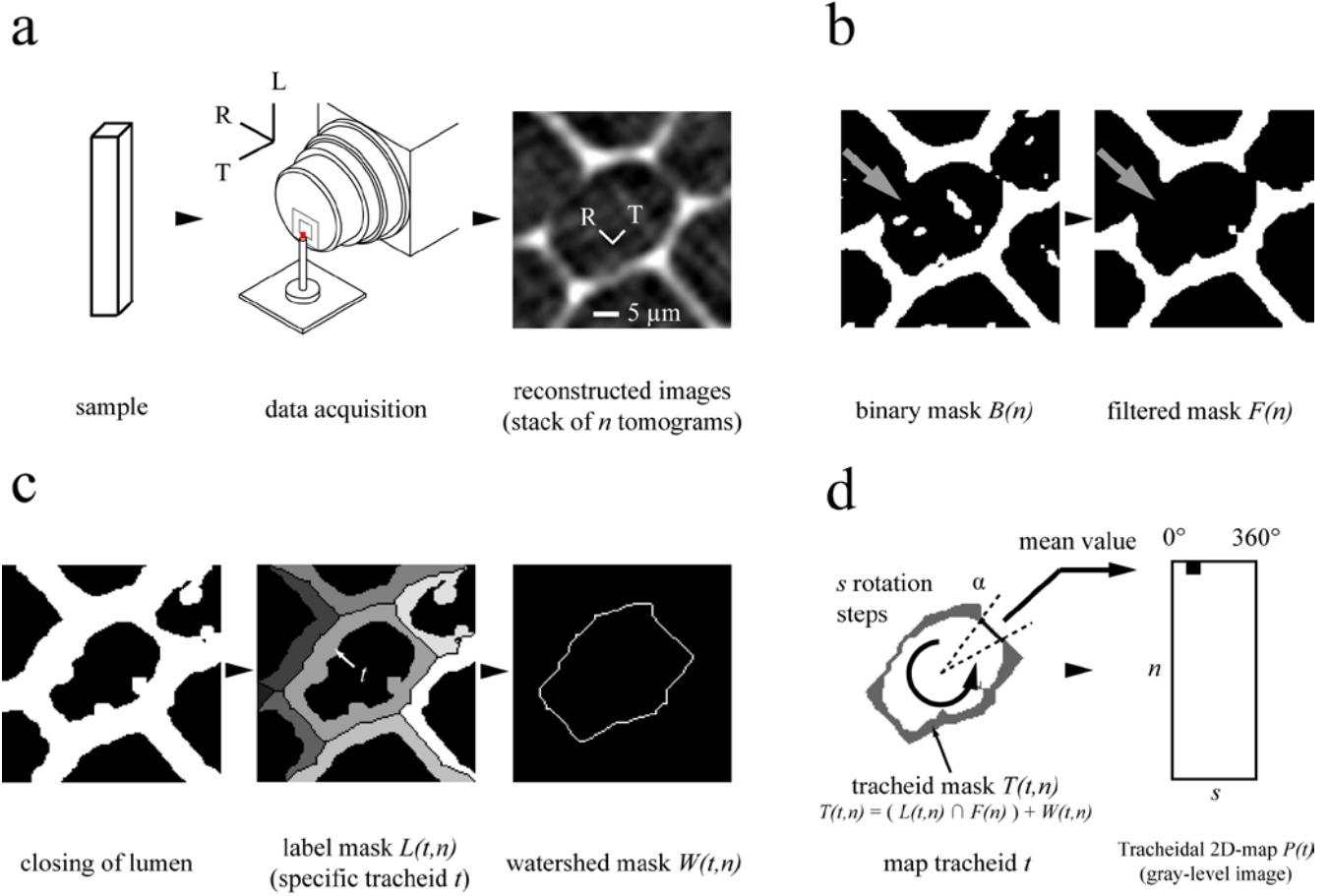

fig1_fullpagewidth_180_125_mm_300dpi.tif

```
1   program Size_Distribution
2   #Read Data
3   I <- Read Stack of n Tomograms
4   Ic <- Crop ROI of I
5
6   #Construction 2D Map
7   for i to n do
8       B <- Compute binary mask of Ic(n)
9       Bc <- Morphological closing of B
10      F(n) <- Remove objects < p pixels from Bc
11  end Return Filtered mask (F)
12
13  Fc <- Morphological closing of F
14  for i to n do
15      S <- Compute skeleton of Fc(n)
16      Sr <- Remove spur pixels from S
17      L(n) <- Compute label mask from Sr and Fc
18  end Return Label mask L
19
20  t <- Choose specific tracheid
21  for i to n do
22      L(t,n) <- Find all pixels of tracheid t in L(n)
23      W(t,n) <- Compute watershed of tacheid t
24      Compute tracheid mask T(t,n) = ( L(t,n) ∩ F(n) ) + W(t,n)
25      P(t,n) <- Normal cylindric projection P(t,n)
26  end Return Normal cylindrical projection P(t)
27
28  #Size distribution
29  Pb(t) <- Construct binary image of P(t)
30  Compute size distribution of cell wall elements in Pb(t)
31  end program Size_Distribution
```

fig2_onehalfcolumn_170_130_mm.eps

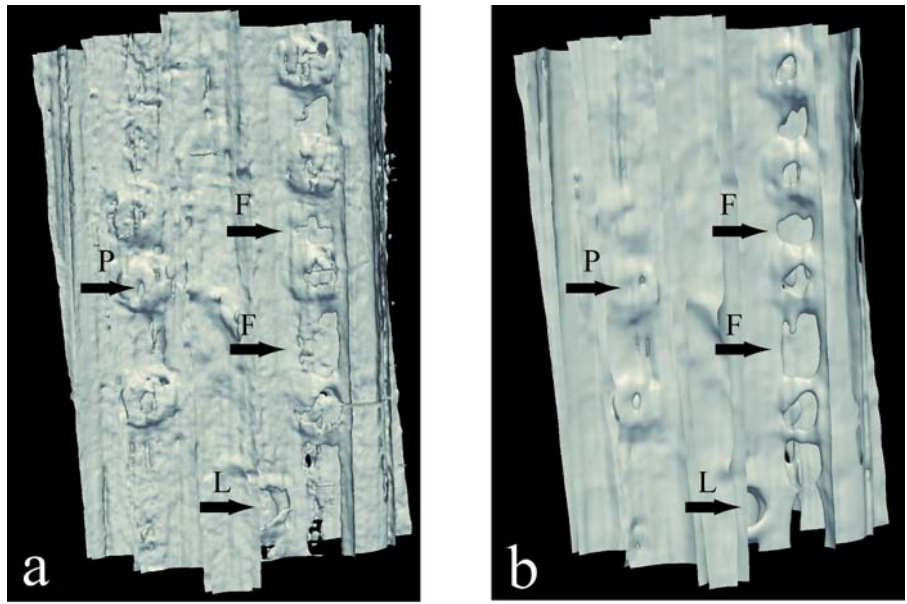

fig3_onehalfcolumn_180_80_mm.tif

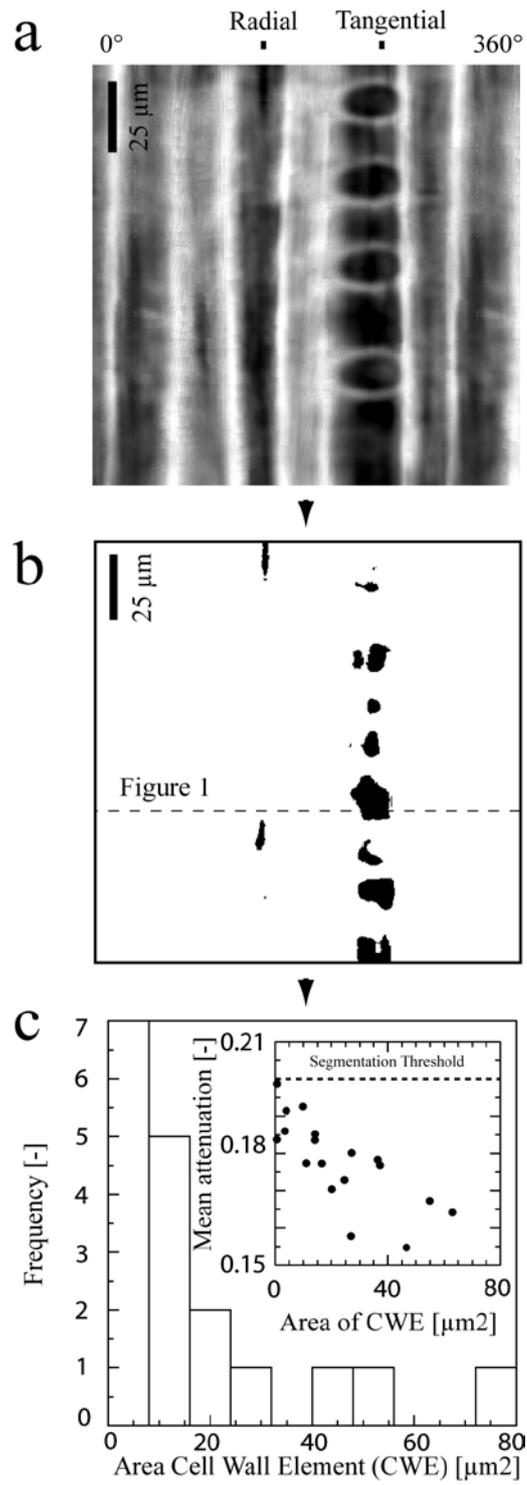

fig4_onecolumne_70_200_mm.tif